\documentclass[traditabstract,letter]{aa}
\usepackage{graphicx}
\usepackage{txfonts}
\usepackage[normalem]{ulem}
\usepackage{placeins} 
\begin{document} 

\title{Discovery of a kinematically distinct component in the central region of the collisional ring galaxy AM0644-741}

\author{Chayan Mondal
          \inst{1,2}\fnmsep\thanks{corresponding author}
          \and
          Sudhanshu Barway\inst{3}
          }

   \institute{Academia Sinica Institute of Astronomy and Astrophysics (ASIAA), No. 1, Section 4, Roosevelt Road, Taipei 106319, Taiwan\\
              \email{cmondal@asiaa.sinica.edu.tw, mondalchayan1991@gmail.com}
              \and
              Inter-University Centre for Astronomy and Astrophysics, Ganeshkhind, Post Bag 4, Pune 411007, India
         \and
             Indian Institute of Astrophysics, Koramangala II Block, Bangalore-560034, India\\
             }

   \date{}

\abstract{We present the discovery of a peculiar central stellar structure in the collisional ring galaxy AM0644-741 using HST imaging and MUSE integral field unit (IFU) data. We identified two S\'ersic components with a S\'ersic index of 1.72 (inner part) and 1.11 (outer part) in the HST F814W band optical image using \textsc{Galfit}. We utilized the MUSE data cube to construct stellar line of sight velocity (V$_{\rm LOS}$), velocity dispersion ($\sigma_{\rm LOS}$), h$_3$ \& h$_4$ velocity moments, and stellar population age maps using the \textsc{GIST} pipeline for further investigating both S\'ersic components, which have a difference of $\sim$ 60 degrees in their position angle. The inner component, with an effective radius $\sim$1 kpc, shows a strong anticorrelation between V$_{\rm LOS}$/$\sigma_{\rm LOS}$ and h$_3$, indicating the presence of a rotating stellar structure. In addition, the inner component also shows a relatively higher velocity dispersion (average values reaching up to $\sim$240 km sec$^{-1}$) along with disky isophotes and stronger Mg~$b$ line strength, which all together highlight a peculiar dynamical state of AM0644-741's central region. Our analysis suggests that the recent encounter has had a smaller impact on the stellar orbits within the inner component. In contrast, it has specifically affected the stellar orbits of the progenitor's outer disk when forming the star-forming ring. The BPT analysis of the unresolved nuclear source shows a LINER-type ionization, hinting at AGN activity in the galaxy. Our study projects the dynamical evolution of collisional systems and provides scope for simulations to explore the central region in greater detail.}

\keywords{galaxies: individual: AM0644-74 -- galaxies: structure -- galaxies: kinematics and dynamics -- galaxies: interactions -- galaxies: active -- galaxies: evolution}

\maketitle

\section{Introduction}
Collisional ring galaxies (CRGs) are known to form through the encounter of two galaxies in which a satellite compact galaxy crashes through the disk of a larger progenitor galaxy \citep{lynds1976, Theys1976, appleton1996}. Such encounters produce radially propagating density waves, resulting in the formation of rings \citep{lynds1976}. The ring hosts the majority of the ongoing star formation in the galaxy, whereas it is generally quenched in the enclosed disk \citep{marston1995, appleton1997, Korchagin2001}. Depending on the intruder galaxy and the nature of the collision, the central region of the progenitor also goes through morphological/dynamical changes \citep{athan1997,renaud2018,elagali2018}. A stellar bar was discovered in the central region of the Cartwheel galaxy, and its pre-existing nature was shown by studying the kinematics and stellar populations in the bar region using integral field spectroscopy data from MUSE \citep{barway2020, Mondal2024}. A photometric and kinematic analysis of the central region of such galaxies is required to understand how a collision can affect the disk and other pre-existing structural components.

AM0644-741 and the Cartwheel galaxy \citep{arp1987,Few1982,marcum1992,appleton1997,Higdon2011} host two of the physically largest rings among the CRGs, and are often used to illustrate the success of collisional ring formation. However, a recent study by \citet{pasha2025} reported an even larger ring of diameter $\sim$ 140 kpc in the galaxy Bullseye that hosts nine ring structures. The ring of the galaxy AM0644-741 (the 'Lindsay-Shapley Ring') has a major axis diameter of $\sim$ 42 kpc \citep{Higdon2011}. The kinematics of the ring suggests that the galaxy has undergone a collision $\sim$ 133 Myr ago, which confirms this encounter to be more recent than Cartwheel's \citep{Higdon2011}. The position of the nucleus with respect to the asymmetrically distributed ring indicates an off-centered collision \citep{antunes2007, Higdon2011}. Optical spectroscopy by \citet{Few1982} showed the ring to be rich in giant star-forming complexes and expanding at a velocity of $\sim$ 128 km/sec in accordance with the collisional scenario of ring formation. Utilizing molecular data from the SEST and the H~I observations from the ATCA, \citet{Higdon2011} found the ring to be predominant in atomic gas ($f_{mol}$ = 0.062) with no evidence of molecular gas in the nuclear region. However, \citet{horellou1995} reported the presence of molecular gas in the central region of AM0644-741.

The majority of the studies on AM0644-741 targeted the star-forming outer ring. The properties of the inner stellar disk, which carries the imprint of the progenitor galaxy, have not been explored well. Considering the morphology and the grand size of the ring, the galaxy AM0644-741 offers an ideal platform to understand the structural evolution of a typical collisional system. In this paper, we study the photometric and kinematic properties of the inner region of the galaxy AM0644-741 to understand its post-collisional nature. The paper is arranged as follows: Section \S\ref{sec:data} presents the details of data used, Section \S\ref{sec:analysis} describes the analysis, the discussion in Section \S\ref{sec:discussion}, followed by the conclusion in \S\ref{sec:conclusion}. We adopt a cosmology with $H_0 = 70$ km s$^{-1}$ Mpc$^{-1}$, $\Omega_{\Lambda} = 0.7$, $\Omega_M = 0.3$\, and provide all magnitudes in the AB system.

\begin{figure}
    \centering
    \includegraphics[width=3.in]{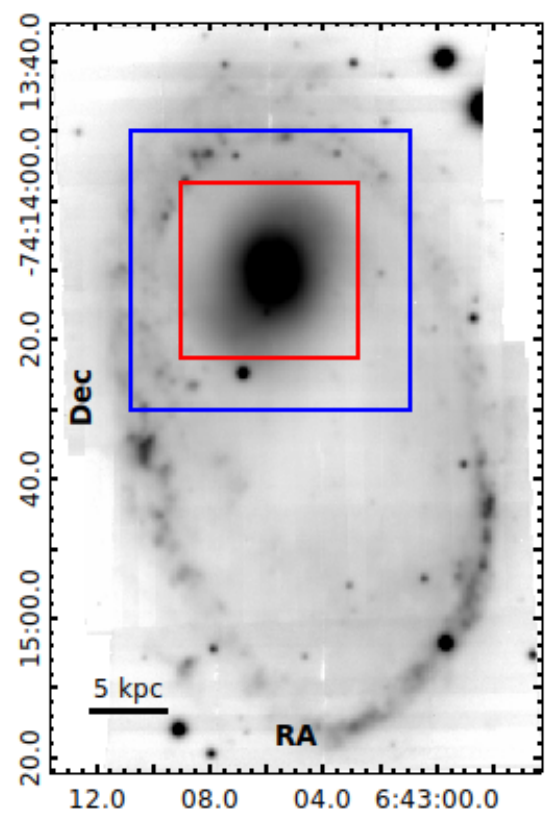} 
    \caption{MUSE white light image of the galaxy AM0644-741 constructed by combining frames between 4800 - 5800 \AA~ from the IFU cube. The red and blue rectangles show the extent of the HST F814W band image and the MUSE data cube analyzed in this study, respectively.}
    \label{fig_full}
\end{figure}

\section{Data}
\label{sec:data}
We used the Integral Field Unit (IFU) spectroscopic data of the galaxy AM0644-741 observed using the Multi-Unit Spectroscopic Explorer (MUSE) instrument installed at ESO's Very Large Telescope (VLT) \citep{bacon2010}. Each spaxel (size$\sim$ 0\farcs2 $\times$ 0\farcs2) of the MUSE IFU (field of view $\sim$ 1$^{\prime} \times 1^{\prime}$) provides an optical spectrum covering a wavelength range 4750 - 9350 \AA{}. The IFU offers a spectral resolution of $\sim$ 1.25 \AA~, enabling deblended identification of many important optical emission lines. We acquired the IFU data of the galaxy AM0644-741 from the ESO science portal (Program ID - 106.2155.001). The entire MUSE deep cube covers a rectangular area of $\sim$ 1.0$^{\prime} \times 1.8^{\prime}$ (Figure \ref{fig_full}). We created a smaller cutout of size $\sim$ 40$^{\prime\prime} \times 40^{\prime\prime}$ (area marked in blue rectangle in Figure \ref{fig_full}) for analyzing only the inner part of AM0644-741 in this study. We note here that by `inner part of the galaxy' or `the galaxy' we would mean the region inside the blue rectangle in the rest of the paper. The seeing of the MUSE observation is reported as $\sim$ 1\farcs16, which allows us to characterize the spectral properties of the reported central source in the galaxy. We also used the optical image (size $\sim$ 25$^{\prime\prime} \times 25^{\prime\prime}$, i.e., the area marked in the red rectangle in Figure \ref{fig_full}) in the F814W filter observed with ACS wide field channel onboard the Hubble Space Telescope (HST) to perform a photometric structural decomposition. The HST image, with $\sim$ 10 times better angular resolution than MUSE, is particularly useful to identify any existing smaller structural components in the central part of the galaxy.

\begin{figure}
    \centering
    \includegraphics[width=3.5in]{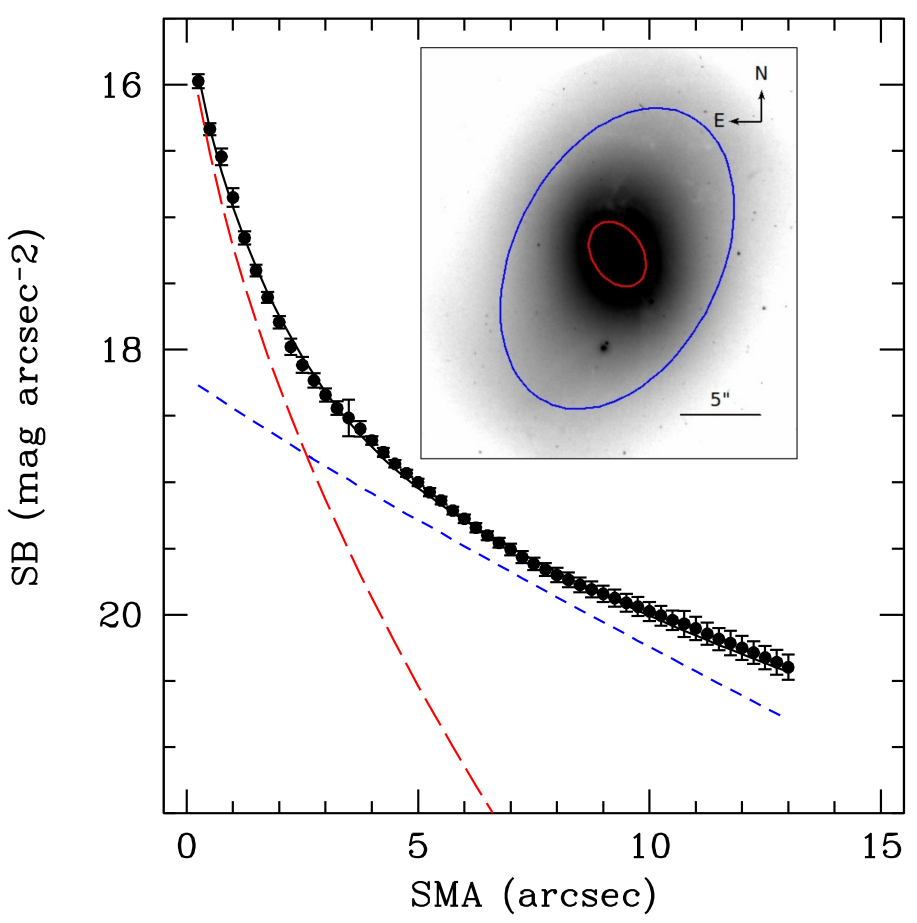}
    
    \caption{Surface brightness profiles derived using the IRAF ELLIPSE task from the HST F814W band observed and \textsc{Galfit} model images. The black points denote the values derived from the observed image, whereas the profiles for the inner S\'ersic, outer S\'ersic, and the combined model are shown in red dashed, blue dashed, and black solid lines, respectively. In the inset, we show the HST image and the extent of both components in respective colors, considering values as specified in Table \ref{table_galfit}. The errors, shown for the observed surface brightness, are calculated from the rms scatter of the isophotal intensity.}
    \label{fig_galfit}
\end{figure}

\section{Analysis}
\label{sec:analysis}

\subsection{Structural components from 2D image decomposition}
\label{sec:galfit}
We used \textsc{Galfit} \citep{peng2002} to perform a 2D decomposition of the inner part (as shown in Figure \ref{fig_galfit}) in the HST F814W band image. A good fit was obtained using two Sérsic functions, while the central object was modeled as an unresolved point source convolved with the PSF, which has a total F814W band magnitude of 19.29 mag. The resulting surface brightness profiles, derived using the IRAF ELLIPSE task, from the observed and model images are shown in Figure \ref{fig_galfit}, and the model-derived parameters are listed in Table \ref{table_galfit}. For better visualization of the \textsc{Galfit} modeling, we have also shown the observed image, best-fit model, and the residual map in Figure \ref{fig_galfit_2d}. The outer component has a S\'ersic index ($n_b$) of 1.11, which signals an exponential disk, whereas detection of the inner S\'ersic component, with $n_b$ = 1.72, could indicate the presence of a bulge-like structure in the galaxy. The effective radii ($r_e$) of the inner and outer components are found to be 2\farcs19 ($\sim$0.97 kpc) and 9\farcs98 ($\sim$4.44 kpc), respectively. We noticed $\sim 60$ degree difference between the position angle (PA) of the two fitted S\'ersic models (clear from the PA profile in Figure \ref{fig_ellipse} as well), which strengthens the existence of two distinct components in the galaxy. The Ellipticity and PA profiles, shown in Figure \ref{fig_ellipse}, exhibit isophotal twisting which further confirms the presence of two components (discussed in Appendix \ref{s_appndx_1}). As the galaxy is known to be moderately inclined in the sky (\textit{i}=57.2$^{\circ}$; \citealt{Few1982}), we further examined the effect of inclination on the detected components and found that the observed isophotal twist is not a result of inclination angle (discussed in Appendix \ref{s_appndx_3}). In this work, we present our analysis using the observed images as the measurement of inclination angle in CRGs, especially from the outer ring, can include significant bias in the case of non-isotropic expansion of the ring (eg., \citealt{bizyaev2007}). As a result, one has to be cautious while interpreting structures even from a deprojected image.

\begin{table*}
\centering
\caption{Results of the \textsc{Galfit} model fitting on the HST F814W band image of the galaxy as shown in Figure \ref{fig_galfit}}
\label{table_galfit}
\begin{tabular}{p{2cm}p{2cm}p{2.5cm}p{1cm}p{1cm}p{1cm}}
\hline
Component & F814W & Effective & S\'ersic & axis & PA\\
 & magnitude & radius ($r_e$) & index ($n_b$) & ratio (b/a) & (degree)\\\hline

Inner (S\'ersic) & 14.15 & 2\farcs19 (0.97 kpc) & 1.72 & 0.71 & 33.4\\

Outer (S\'ersic) & 12.97 & 9\farcs98 (4.44 kpc) & 1.11 & 0.65 & $-$26.5\\
PSF & 19.29 & - & - & - & -\\

\hline
\end{tabular}
\end{table*}

\begin{figure*}
    \centering
    \includegraphics[width=7in]{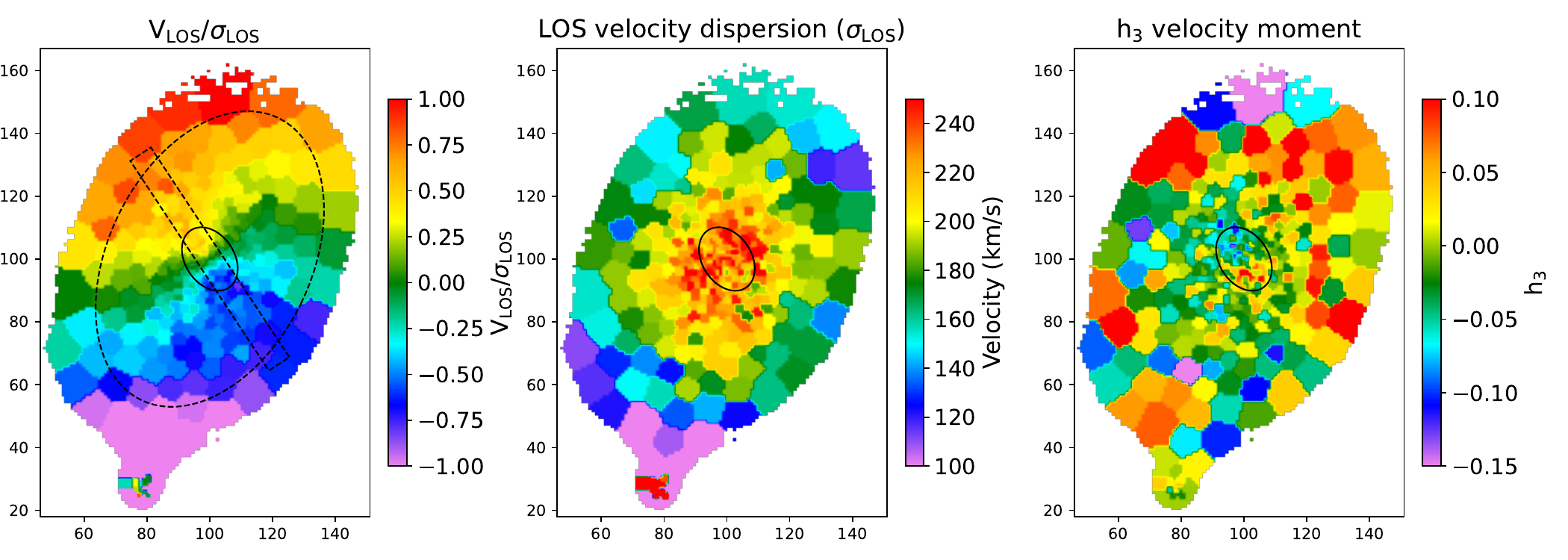} 
    \caption{Stellar kinematic maps of AM0644-741 derived from the MUSE IFU data using the GIST pipeline. The figure shows stellar V$_{\rm LOS}$/$\sigma_{\rm LOS}$ (\textit{left)}, velocity dispersion ($\sigma_{\rm LOS}$) (\textit{middle)}, and h$_3$ velocity moment (\textit{right}) maps of the inner part (i.e., region within the blue rectangle shown in Figure \ref{fig_full}). The ellipses shown in black solid and dashed lines represent the inner and outer S\'ersic components identified in the HST optical image (Figure \ref{fig_galfit}). The black dashed rectangle shows the pseudo slit that was used to derive the radial V$_{\rm LOS}$/$\sigma_{\rm LOS}$ and $\sigma_{\rm LOS}$ profiles (shown in Figure \ref{fig_all_profile}) along the major axis of the detected inner component.}
    \label{fig_kin_maps}
\end{figure*}

\begin{figure*}
    \centering
    \includegraphics[width=7in]{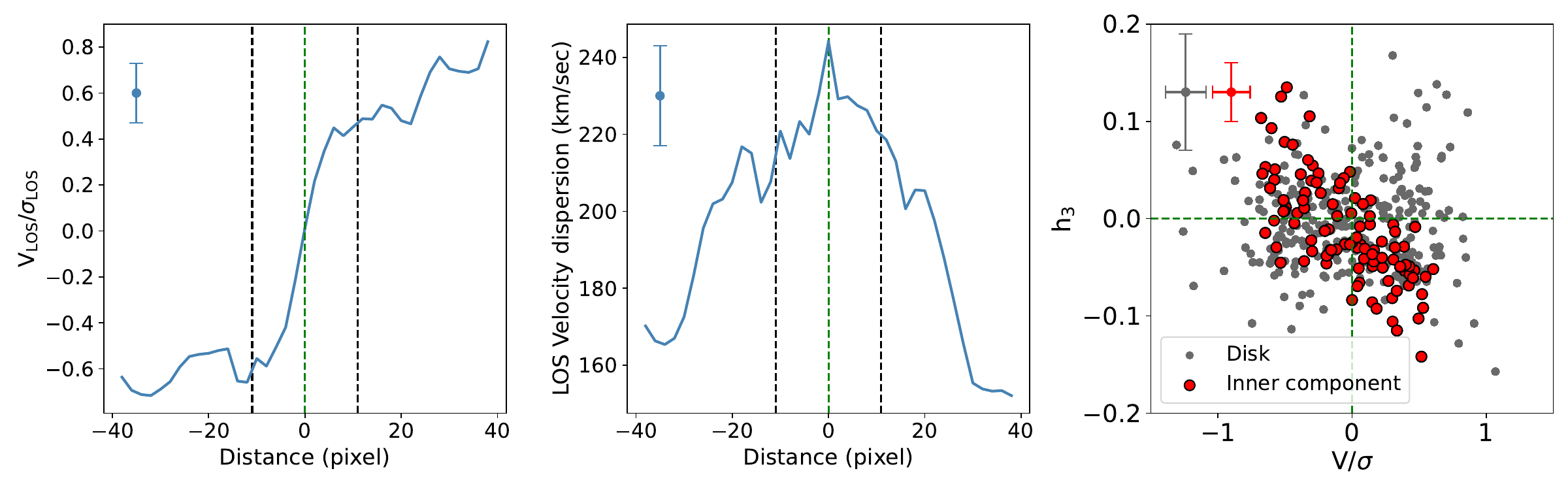} 
    \caption{V$_{\rm LOS}$/$\sigma_{\rm LOS}$ (\textit{left}) and LOS velocity dispersion (\textit{middle}) profiles, measured along the pseudo slit shown in Figure \ref{fig_kin_maps}. The vertical black dashed lines in both panels mark the edge of the identified inner component on either side of the minor axis, which is shown in a green dashed line. The red points on the \textit{right} panel show the V$_{\rm LOS}$/$\sigma_{\rm LOS}$ and h$_3$ values of the spaxels enclosed within the extent of the identified inner component (as shown by the ellipse in solid black line in Figure \ref{fig_kin_maps}). The gray points represent spaxels located outside the inner component. The error bars shown in each panel represent the typical error values associated with each respective quantity.}
    \label{fig_all_profile}
\end{figure*}

\subsection{MUSE kinematics of the detected components}
\label{sec:muse}
We employed the Penalised Pixel Fitting (pPXF) module \citep{cappellari2004,cappellari2017} of the GIST IFU data analysis pipeline \citep{bittner2019} to produce kinematic maps of the galaxy using the MUSE data cube. The observed MUSE data is binned spatially considering the Voronoi binning method \citep{cappellari2003} to improve the signal-to-noise ratio of the spectra before performing the fitting. As a convenient choice, we fixed the upper limit of the S/N as 30 and the lower limit as 3 for the Voronoi binning. The integrated spectrum of each binned unit is fitted with MILES template spectra \citep{vazdekis2010} combined with a Line of Sight Velocity Distribution (LOSVD) function within the wavelength range 4800 - 5800 \AA{} as suggested in \citet{bittner2021,bittner2019} to extract the kinematic maps. As a product of the best-fit solution, we derived the values of stellar LOS velocity (V$_{\rm LOS}$), velocity dispersion ($\sigma_{\rm LOS}$), and higher order velocity moments h$_3$ and h$_4$. In Figure \ref{fig_kin_maps}, we show the spatial maps of V$_{\rm LOS}$/$\sigma_{\rm LOS}$, $\sigma_{\rm LOS}$, and h$_3$ velocity moment. We have shown the LOS velocity and the h$_4$ velocity moment maps in Figure \ref{fig_v_h4}. Both LOS velocity and V$_{\rm LOS}$/$\sigma_{\rm LOS}$ maps show the signature of a rotating stellar disk. We also noticed that the position angle of the line of nodes, observed in the V$_{\rm LOS}$/$\sigma_{\rm LOS}$ map, shows a significant shift from the inner to the outer part, which matches well with the change in the position angle of two Sérsic components. Our findings convey that although the photometric and kinematic axes of each respective Sérsic component are closely aligned, the axes of the inner component are misaligned by $\sim$60 degrees with respect to that of the outer component, which reinforces the existence of two distinct units in the galaxy. Furthermore, the overall elevated h$_4$ values, observed within the extent of the inner component (Figure \ref{fig_v_h4}), suggest the presence of two superimposed structural components in the central part of the galaxy \citep{gadotti2020}. The V$_{\rm LOS}$/$\sigma_{\rm LOS}$ and h$_3$ show a clear trend of anticorrelation within the extent of the inner component. However, the LOS stellar velocity dispersion in the inner part is relatively higher ($\sigma \sim 200 - 260$ km/s) than the typical values observed in normal disk galaxies \citep{mogotsi2019, ohsree2020, gadotti2020}. Considering the inclination angle of the galaxy, we note here that the actual values of velocity and velocity dispersion would be higher than the derived LOS measurements. However, the trend observed in different maps shown in Figure \ref{fig_kin_maps}, which are primarily utilized for interpreting the kinematic nature of the inner region, will persist even after correcting for the inclination effect.

\section{Discussion}
\label{sec:discussion}
\subsection{Photometric and kinematic nature of the inner component}
\label{sec:bulge_nd}

Collisional ring galaxies are ideal laboratories for studying the impact of galaxy interaction on their structural evolution and the subsequent star formation. The newly formed rings, which host sites of intense star formation in the post-collisional phase, carry important clues about the dynamics of the collision. The majority of the studies on CRGs have targeted these star-forming rings, keeping the central region, which carries the imprint of the progenitor's disk, less explored. Our findings in the galaxy AM0644-741 are therefore important to understand the evolution of the underlying disk in CRGs. Our photometric image decomposition has identified two Sersic components in the galaxy with S\'ersic index values 1.72 (inner) and 1.11 (outer), signifying the presence of two distinct stellar structures. To better understand the photometric nature of these structures, we have inspected the ellipticity, position angle, and Fourier component b$_4$ profiles of the fitted isophotes (Figure \ref{fig_ellipse}). The Ellipticity and PA profiles strengthen the presence of two distinct components, while the values of b$_4$ signify the inner component to be disky and the outer component to have a mixed nature (discussed in Appendix \ref{s_appndx_1}).

The LOS stellar kinematic maps also show promising evidence of two distinct structures in the galaxy. The V$_{\rm LOS}$/$\sigma_{\rm LOS}$ map shows a clear signature of rotation, which is stronger in the inner part (Figure \ref{fig_kin_maps}). This is also clear in the radial V$_{\rm LOS}$/$\sigma_{\rm LOS}$ profile (Figure \ref{fig_all_profile} - left panel), which shows a steeper gradient within the extent of the inner component. To further explore the kinematic nature of both components, we plot the V$_{\rm LOS}$/$\sigma_{\rm LOS}$ and h$_3$ values of all the spaxels belonging to each respective component in Figure \ref{fig_all_profile} (right panel). The spaxels part of the inner component (i.e., those within the black solid ellipse in Figure \ref{fig_kin_maps}) shows a stronger trend of anticorrelation (Spearman coefficient = $-0.75$) between V$_{\rm LOS}$/$\sigma_{\rm LOS}$ and h$_3$ (red points in Figure \ref{fig_all_profile} - right panel), which is a signature of near-circular stellar orbits observed in a stable galaxy disk \citep{bender1994,bittner2019,gadotti2020}. Though we see a disk-like rotating body in the central part, the LOS velocity dispersion is found to be relatively higher within the same region (Figure \ref{fig_kin_maps}). We employed a pseudo slit along the major axis of the inner component (Figure \ref{fig_kin_maps}) and derived the radial velocity dispersion profile (Figure \ref{fig_all_profile}) that shows a central peak with average values reaching $\sim$ 240 km/s, which is around 3 times higher than the values observed in Cartwheel \citep{Mondal2024}. Such velocity dispersion values with characteristic central peaks are more common in stellar bulges than normal disks \citep{fisher2016}.

Therefore, the galaxy AM0644-741 hosts two components, among which the outer one plausibly represents an exponential disk of effective radius 4.4 kpc, whereas the central component of r$_e\sim$ 1 kpc shows a stronger rotational signature, disky isophotes (i.e., b$_4>0$) along with a relatively higher velocity dispersion. The stronger rotational support observed within the inner component could also indicate the central part of the compact stellar disk that the progenitor galaxy had before the encounter. Considering the predominance of older stellar populations of age $\sim$11 Gyr in the inner part (see discussion in Section \ref{s_appndx_2} and Figure \ref{fig_age_map}), we infer that the detected rotating stellar structure has most likely been formed in the progenitor galaxy and the stellar orbits within it have not been affected significantly due to the recent encounter that the galaxy AM0644-741 experienced around 133 Myr ago. From the presence of a relatively smaller amount of massive star formation in AM0644-741 compared to the Cartwheel, \citet{Higdon1997} indicated that the progenitor of AM0644-741 had a thicker pre-collisional disk. This further justifies the existence of a post-collisional disk-like structure in the galaxy. In the case of the outer component, the weaker anticorrelation between V/$\sigma$ and h$_3$ and the presence of both boxy and disky isophotes (discussed in Appendix \ref{s_appndx_1}) signify that the collision has impacted the stellar orbits in the outer part of the progenitor's disk more strongly while forming the star-forming ring.

\subsection{Characterizing the central region using optical emission and absorption lines}

To improve our understanding of the central region, we explored the properties of identified emission and absorption lines in the MUSE optical spectrum. The integrated spectrum extracted from spaxels within the inner component shows the presence of Mg~$b$ absorption line (see Appendix \ref{s_appndx_4} \& Figure \ref{fig_nuclear}). Following the method discussed in \citet{kunt2006}, we estimated the Mg~$b$ line index as 4.9 \AA{}, which falls within the range typically seen in elliptical galaxies and/or classical bulges \citep{dressler1987,ziegler1997,fisher2016}. The stronger Mg~$b$ line index justifies the relatively higher velocity dispersion observed in the inner region of the galaxy, suggesting the plausible emergence of a dynamically hotter central structure. 

\citet{wolter2019} reported the detection of an unresolved central source in X-ray with a luminosity L$^{[0.5 - 10~keV]}_{X}$ = 22.4$\times~10^{39}$ erg s$^{-1}$, which is higher than the typical values observed in Ultra Luminous Xray sources. We showed the BPT measurement of the nuclear source in Figure \ref{fig_nuclear} (described in Appendix \ref{s_appndx_4}), which indicates a LINER-type ionization. This further justifies the absence of star formation and molecular gas in the nuclear region as reported by \citep{Higdon2011}. This is also true in general for CRGs, which show a reduction in ongoing star formation in the inner part \citep{romano2008,ahmed2018}. Our measurements, combined with the nature of X-ray emission from the nuclear source, signify the presence of an AGN in the galaxy AM0644-741. Including the peculiar behavior noticed in stellar kinematics, the detection of AGN activity makes the inner part of the galaxy even more compelling. While the AGN could also have a pre-collisional origin, we speculate that the recent collision might have played a crucial role in triggering AGN activity and also in gearing up the stellar velocity dispersion in the central region of the galaxy.  

\section{Conclusion}
\label{sec:conclusion}

Our photometric and spectroscopic analysis of the galaxy AM0644-741 has identified two stellar structures of different properties. The inner component of $r_e$ $\sim$ 1 kpc shows: disky isophotes ($b_4>0$), strong stellar rotational support (V$_{\rm LOS}$/$\sigma_{\rm LOS}$ - h$_3$ anticorrelation), relatively higher velocity dispersion with a central peak ($\sigma_{\rm LOS}\sim$240 km sec$^{-1}$), and stronger Mg~$b$ line index. Combining all the evidence, we conclude that the central stellar structure in AM0644-741 has a peculiar dynamical state with reference to different known secularly built structures. While the observed trend of stellar rotation implies a minor impact of the recent collision on stellar orbits within the inner part, the elevated value of velocity dispersion and Mg~$b$ line index could also indicate some level of dynamical heating in the central region. The other component, with a difference of $\sim60$ degrees in PA, has a more disk-like shape ($n_b = 1.11$) of r$_e$ = 4.44 kpc. However, the isophotes part of this extended component has both boxy and disky shapes along with weaker V$_{\rm LOS}$/$\sigma_{\rm LOS}$ - h$_3$ anticorrelation, which plausibly suggests that the recent encounter have impacted the stellar orbits in progenitor's outer disk more strongly. The mass-weighted age of the stellar populations indicates that both these structures have a pre-collisional origin. Our study conveys that the dynamical evolution of post-collisional systems can be unique and needs more attention to explore with detailed simulations. We also identified a LINER-type ionization in the nuclear region through a BPT diagnostic plot. Though a low-luminosity AGN could have already existed in the progenitor galaxy, we speculate that the recent encounter might have enhanced AGN activity in the galaxy.

\begin{acknowledgements}
We thank the anonymous referee for valuable suggestions. This paper made use of Matplotlib \citep{matplotlib2007}, Astropy \citep{astropy2013,astropy2018}, and SAOImageDS9 \citep{joye2003}. This work has used observations collected at the European Southern Observatory under ESO program 106.2155.001. This research has made use of the NASA/IPAC Extragalactic Database (NED), which is operated by the Jet Propulsion Laboratory, California Institute of Technology (Caltech), under contract with NASA. This research is based on observations made with the NASA/ESA Hubble Space Telescope obtained from the Space Telescope Science Institute. We are grateful to Adrian Bittner for his numerous suggestions, which helped us to understand the GIST pipeline used in this work. 
\end{acknowledgements}


\begin{appendix}
\onecolumn

\section{Additional photometric and kinematic analysis}
\label{s_appndx_1}
\begin{figure*}[h!]
    \centering
    \includegraphics[width=7in]{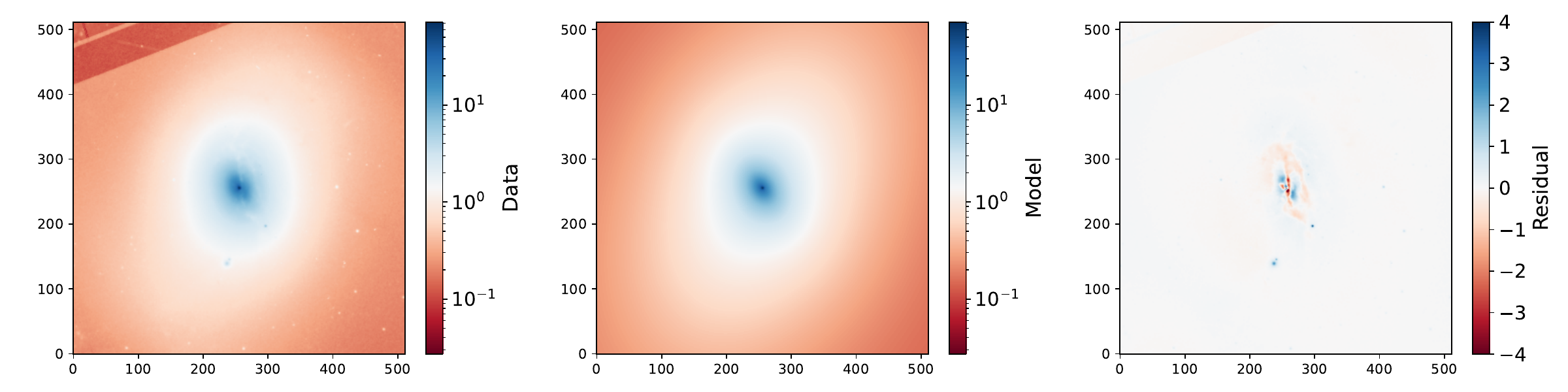}
    \caption{The 2D images to highlight the \textsc{Galfit} modeling of the detected components. The figure shows the observed HST F814W band image \textit{(left)}, model image \textit{(middle)}, and the residual map \textit{(right)}.}
    \label{fig_galfit_2d}
\end{figure*}

In this appendix, we provide some relevant additional results to support our inferences presented in the paper. In Figure \ref{fig_galfit}, we have shown the surface brightness profile of the galaxy, derived using IRAF ELLIPSE task, from the HST F814W band observed and model (generated using \textsc{Galfit}) images. However, to better visualize the \textsc{Galfit} image decomposition, we show the 2D observed and model images along with the residual map in Figure \ref{fig_galfit_2d}. The goodness of the fit can be seen from the overall small residuals. The filamentary vertical structure in the central region of the residual map originates due to the dust lanes, which can also be seen in the actual image of the galaxy.

\begin{figure*}[h!]
    \centering
    \includegraphics[width=6.5in]{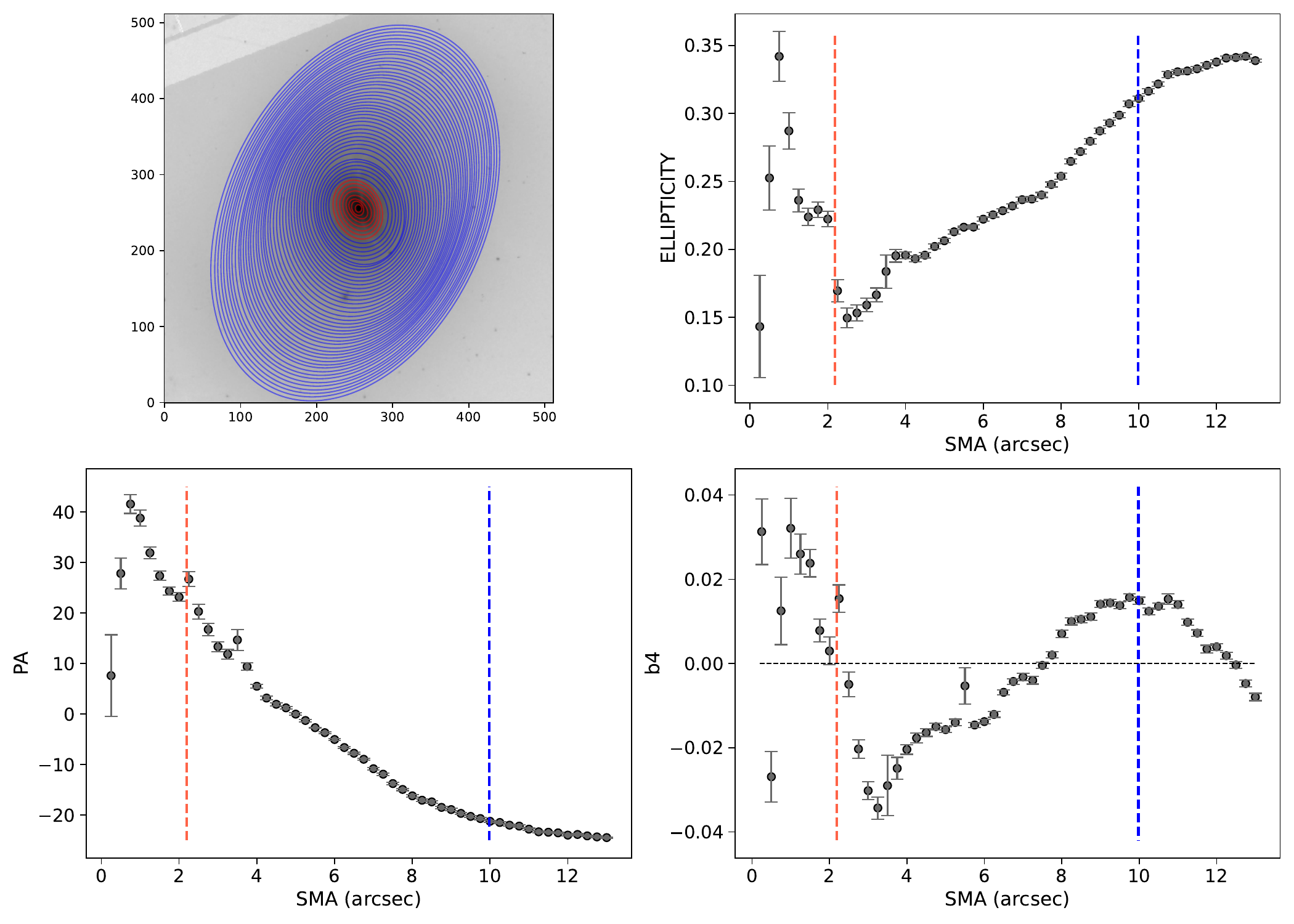}
   
    \caption{Results derived using the IRAF ELLIPSE package on the HST F814W band image of the galaxy. The fitted ellipses are shown in blue on the HST F814W band image \textit{(top left)}. The ellipticity, position angle, and the Fourier coefficient b$_4$ of the fitted ellipses with the length of their semi-major axes are shown in the \textit{top right}, \textit{bottom left}, and \textit{bottom right} panels, respectively. The vertical red and blue dashed lines, shown in these three profiles, mark the extent of the detected inner and outer components, respectively.}
    \label{fig_ellipse}
\end{figure*}

We have shown the extent and projected shape of both detected components in the inset of Figure \ref{fig_galfit}. However, to inspect the shape better, we further show the fitted isophotal ellipses in Figure \ref{fig_ellipse} (top left) along with their ellipticity (top right), position angle (bottom left), and Fourier coefficient b$_4$ (bottom right) profiles. The ellipticity (i.e., 1 $-$ b/a) and position angle highlight the observed projected shape of the isophotes, whereas the value of b$_4$ can infer the disky (b$_4>0$) or boxy (b$_4<0$) nature of the fitted component \citep{jedrze1987}. The ellipticity values of the isophotes, located within the extent of the inner component, overlap with the overall range ($\sim$ 0.14 - 0.35) observed in the outer component, indicating no significant difference in the projected shape of both components. This also agrees with the similar axis ratio (i.e., b/a) values estimated for both S\'ersic components (Table \ref{table_galfit}). However, the ellipticity profile does show a transition near the boundary of the inner component. We also notice a gradual change in the PA angle of the fitted ellipses from center to outer part, reinforcing the presence of two distinct structures that we have detected using \textsc{Galfit}. Interestingly, the majority of the isophotes within the inner component have positive $b_4$ values, signifying their disky nature. The isophotes, around the boundary of the detected outer component, also have positive b$_4$ values, indicating disky signature. However, the isophotes from the edge of the inner component up to $\sim7\farcs6$ within the outer component show a boxy nature (i.e., b$_4<$0). The transition of b$_4$, within the span of the outer component, could indicate the impact of the collision on the progenitor's main disk, adding more peculiarity to post-collisional systems.

\begin{figure*}[h!]
    \centering
    \includegraphics[width=6.5in]{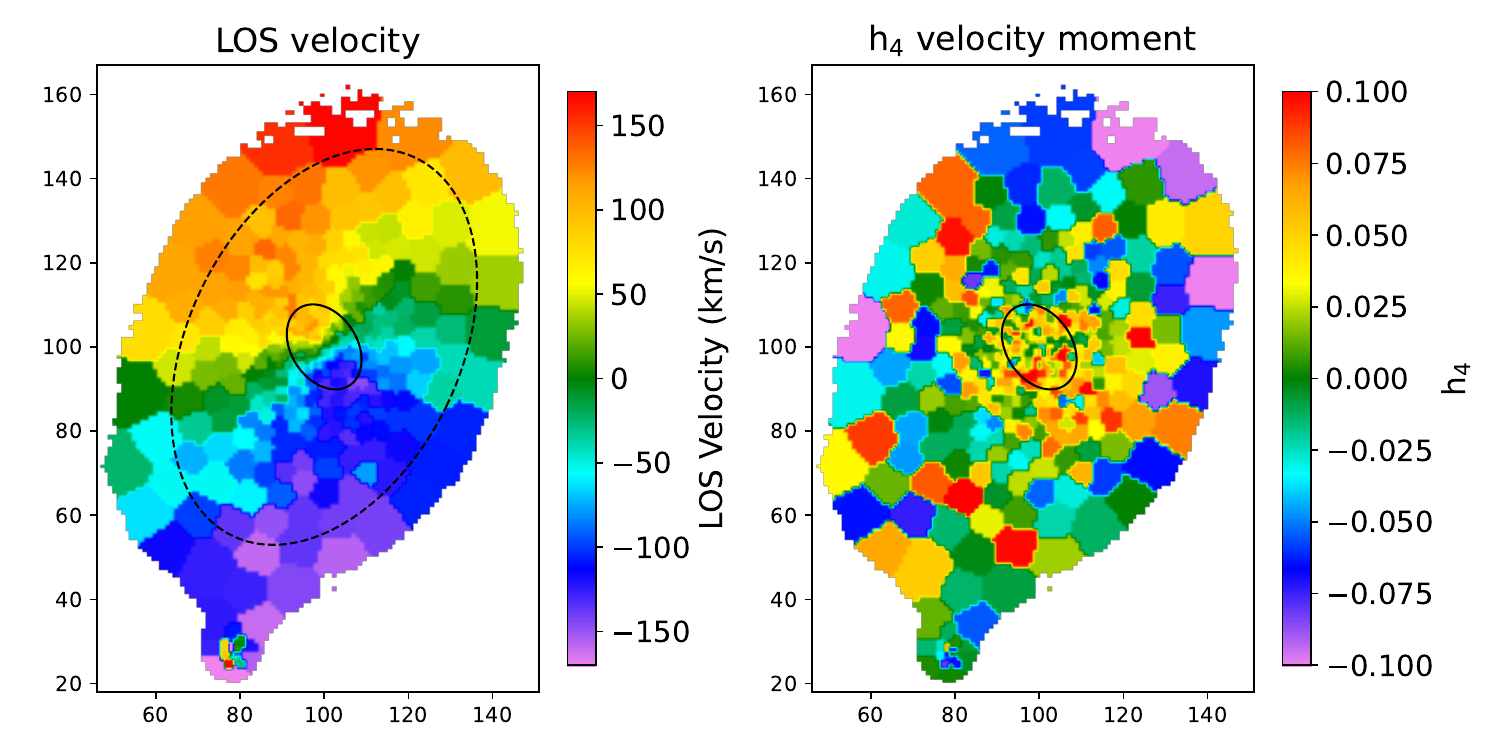}
    \caption{Stellar LOS velocity and h$_4$ velocity moment maps of AM0644-741 for the same region as shown in Figure \ref{fig_kin_maps}. The ellipses shown in solid and dashed black lines are also the same as displayed in Figure \ref{fig_kin_maps}.}
    \label{fig_v_h4}
\end{figure*}

In Figure \ref{fig_v_h4}, we have shown the LOS velocity and h$_4$ velocity moment maps produced from the MUSE data cube. The rotational signature, discussed with reference to the V$_{\rm LOS}$/$\sigma_{\rm LOS}$ map (Figure \ref{fig_kin_maps}), can also be seen in the LOS velocity map of the galaxy. The position angle shift of the line of nodes from the inner to the outer part is also evident.

\FloatBarrier

\section{Effect of Inclination angle}
\label{s_appndx_3}

To understand the effect of inclination angle on the observed isophotes, we performed isophotal ellipse fitting analysis on the deprojected galaxy image. Using the observed orientation of the outer ring, \citet{Few1982} estimated an inclination angle of 57.2 degrees for AM0644-741. We derived the position angle of the ring as 7.7 degrees from the HST image (as it is not provided by \citet{Few1982}). We also used the \textsc{Galfit} measurements of the detected outer component from Table \ref{table_galfit} and estimated inclination and position angle as 49.4 and -26.5 degrees, respectively. Considering both these orientations, we deprojected the HST image using IRAF geotran and performed isophotal ellipse fitting. In both cases, we found the position angle of the fitted isophotes to show a gradual change of around 25 degrees from the edge of the inner component to the outer part. This exercise, therefore, confirms that though inclination angle has some effects on the projected shape of the isophotes, the photometric detection of two distinct components in our study is real and not a result of galaxy inclination.

\section{Stellar population age}
\label{s_appndx_2}
\begin{figure*}[h!]
    \centering
    \includegraphics[width=6.5in]{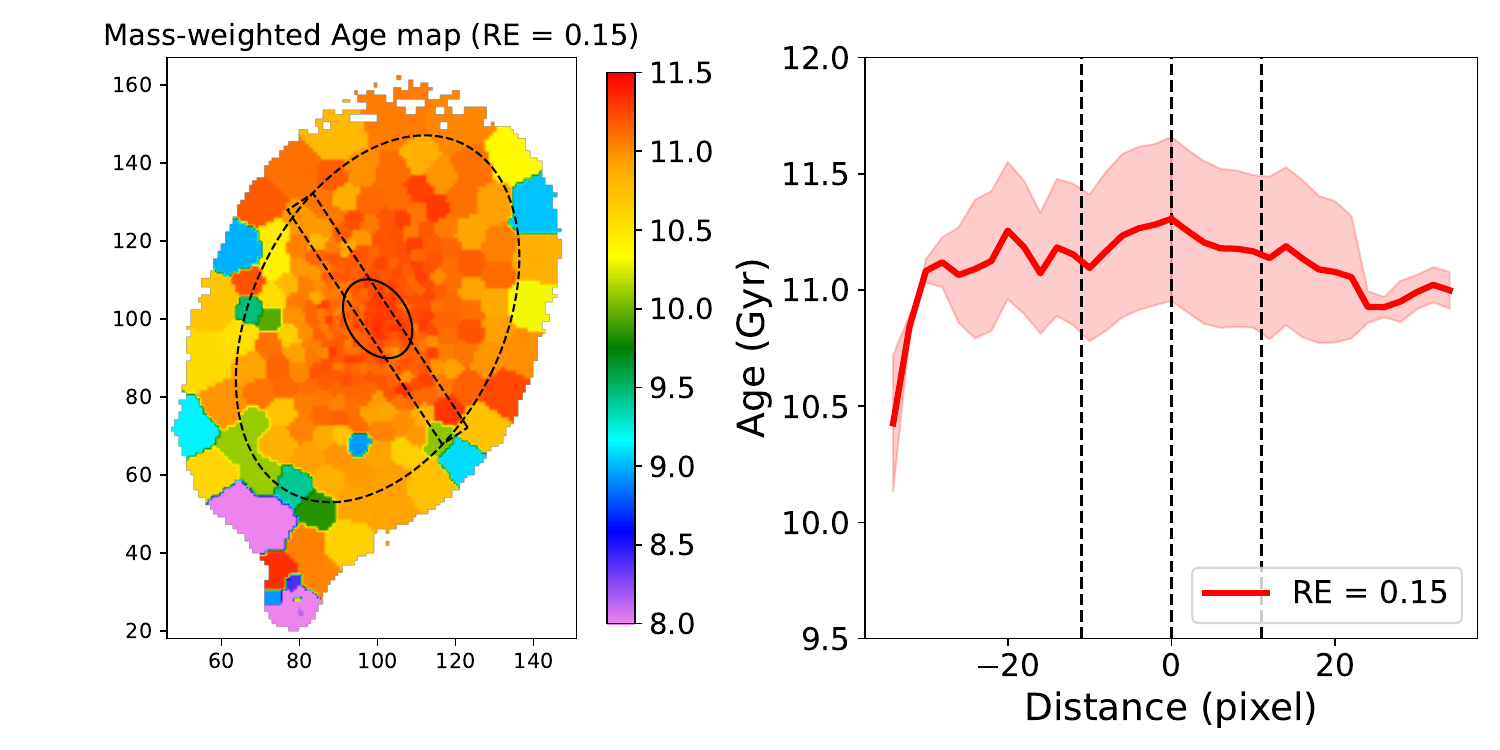}
    \caption{The mass-weighted age map of the stellar populations in AM0644-741 derived from the MUSE data using the GIST pipeline. The \textit{left} panel shows the age map of the same region as displayed in Figure \ref{fig_kin_maps}, estimated with a regularization error (RE) of 0.15. The solid red line on the \textit{right} panel shows the average age values plotted along the pseudo slit as shown on the left panel. The shaded red region around the solid line highlights 1$\sigma$ deviation on the age values estimated considering four different RE values (i.e., RE = 0.10, 0.15, 0.30, 0.45).}
    \label{fig_age_map}
\end{figure*}

We have derived the mass-weighted age map of the galaxy's stellar populations from the same MUSE IFU data using the GIST pipeline. GIST utilizes the regularized pPXF technique to fit the observed continuum using model templates and derive the best-fit age and metallicity of the underlying stellar populations for each Voronoi bin \citep{bittner2019}. We have produced age maps considering four different orders of regularization by choosing REGUL\_ERR as 0.10, 0.15, 0.30, and 0.45 (where regularization = 1/REGUL\_ERR). The age map for REGUL\_ERR = 0.15 (the value adopted throughout the analysis) is shown in Figure \ref{fig_age_map}. The average age profile along the pseudo slit (placed along the major axis of the inner component) is also shown in the same figure.

The mass-weighted age values show that the galaxy's inner part is dominated by older stellar populations of age $\sim$11 Gyr. The structural components identified in the HST optical image, as well as in the MUSE stellar kinematic maps, represent these older populations whose formation has no connection with the recent collision. Therefore, we think the inner structure most likely has a pre-collisional origin and has survived the recent collision. We also note that an N-body dynamical simulation of the entire system would be ideal for understanding the evolution of such unique structural components in a CRG.

\section{Spectroscopic measurement of the central region}
\label{s_appndx_4}
\begin{figure*}[h!]
    \centering
    \includegraphics[width=7in]{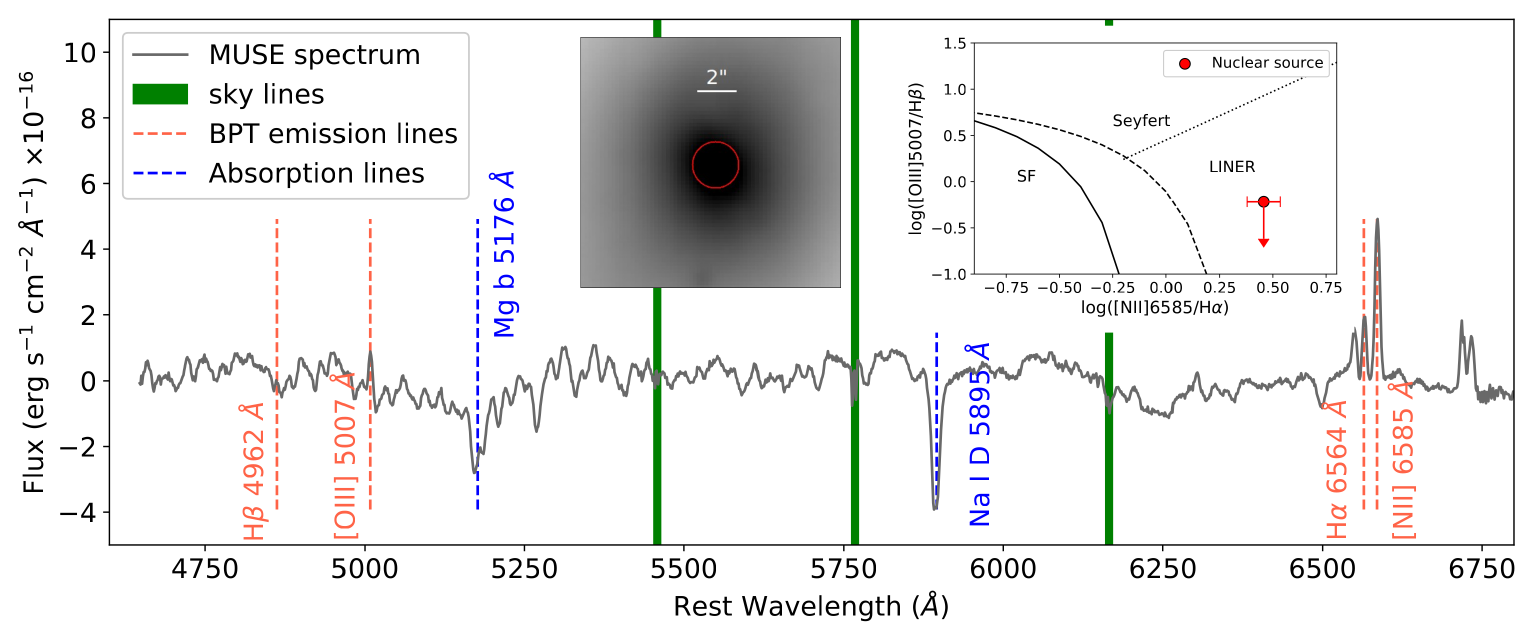}
    \caption{The bluer half of the MUSE spectrum of the nuclear source is shown in gray. The signal is integrated from all the spaxels enclosed within the red circle, as shown in the inset image. The vertical red dashed lines mark four emission lines used to characterize the ionization type through a [NII] BPT diagram, which is also shown in the inset. We showed three diagnostic lines from \citet{kauffmann2003} (solid), \citet{kewley2001} (dashed), and \citet{schawinski2007} (dotted) in the BPT diagram for characterizing our measurements. The vertical green shaded lines represent regions of known optical sky lines. Two prominent absorption lines, Mg~$b$ 5176 \AA{} and Na I D 5895 \AA{}, are marked by blue dashed lines.}
    \label{fig_nuclear}
\end{figure*}

We placed a circular aperture of radius 1\farcs2 ($\sim$ seeing of the MUSE observation) and extracted the integrated MUSE spectrum (Figure \ref{fig_nuclear}) from all the spaxels inside the aperture. We used \textit{lmfit} module and derived continuum-subtracted line flux for [OIII] 5007 \AA{}, [NII] 6583 \AA{}, and H$\alpha$ emission lines to construct the Baldwin, Phillips \& Terlevich (BPT) diagnostic diagram (Figure \ref{fig_nuclear}). As the H$\beta$ line is not detected with a good SNR, we considered the continuum flux at $\lambda$ = 4862 \AA{} to derive an upper limit for [OIII]~5007/H$\beta$ line ratio. In the case of Mg~b~5176 line, we similarly extracted a spectrum integrated from all the spaxels within the identified inner component and estimated the line index following the method described in \citet{kunt2006}.

\end{appendix}

\end{document}